\documentclass[fleqn,12pt,twoside]{article}
\usepackage{espcrc1}

\usepackage{epsfig}

\def\lesssim{\mathrel{\hbox{\rlap{\hbox{\lower4pt\hbox{$\sim$}}}\hbox{$<$}}}}
\def\gtrsim{\mathrel{\hbox{\rlap{\hbox{\lower4pt\hbox{$\sim$}}}\hbox{$>$}}}}

\title{Cosmic Ray Production of $^6$Li by Structure Formation Shocks in the Early Galaxy}

\author{Susumu Inoue
        \address{Max-Planck-Institut f\"ur Astrophysik, Postfach 1317,
                 Karl-Schwarzschild-Str. 1, 85741 Garching, Germany}
        \address[NAO]{National Astronomical Observatory,
                 2-21-1 Osawa, Mitaka, Tokyo, Japan 188-8588}
        and
        Takeru Ken Suzuki
        \addressmark[NAO]
        \address{Dept. of Astronomy, Faculty of Science, University of Tokyo,
                 7-3-1 Hongo, Bunkyo-ku, Tokyo, Japan 113-0033}
        }

\begin{document}

\maketitle

\begin{abstract}
We discuss the production of the element $^6$Li in the early Galaxy
 by cosmic rays accelerated at structure formation shocks,
 driven by the hierarchical merging of sub-Galactic halos during Galaxy formation.
The salient features of this scenario are discussed
 and compared with observations of $^6$Li in metal-poor halo stars,
 including a recent Subaru HDS result on the star HD140283.
Some unique predictions of the model
 are clearly testable by future observations
 and may also provide important insight into how the Galaxy formed.
\end{abstract}

\section{Introduction}
The origin of the light element isotope $^6$Li in old, metal-poor halo stars (MPHS)
 is currently mysterious
 (see \cite{vca00,hob00,rlk00,nis00} for reviews).
Although the most widely discussed models of light element production
 based on nuclear reactions by cosmic rays (CRs) from supernovae (SNe)
 give a good account of the Be and B observed in such stars,
 they generally fall short for the observed $^6$Li.
Such models must resort to rather implausible or contrived assumptions for $^6$Li,
 e.g. a CR injection efficiency substantially higher than normally inferred \cite{rslk00,sy01}
 or an additional low energy CR component lacking observational support \cite{van99}.
We have recently proposed
 a fundamentally different $^6$Li production scenario:
 nuclear reactions induced by CRs accelerated at structure formation (SF) shocks,
 i.e. gravitational virialization shocks driven by the infall and merging of sub-Galactic halos
 during the hierarchical build-up of structure in the early Galaxy \cite{si02}.
Given below is a very brief description of the scenario,
 along with a discussion in relation to recent and future observations.
See \cite{si02,aok02} for more details.

\section{Structure formation cosmic ray scenario: advantages and implications}

SF shocks are inevitable consequences
 of the currently standard theory of hierarchical structure formation in the universe.
The specific energy dissipated at the main SF shock
 accompanying the `final major merger' at redshift $z_f \simeq 2$
 can be evaluated from the expected post-merger virial temperature $T_v$ as
 $\epsilon_{SF} = 3k T_v/2 \simeq 0.4 {\rm keV} (M_t/3 \times 10^{12} {\rm M_\odot})^{2/3} f_c (1+z_f)/3$
 per particle, where $f_c$ is a cosmology dependent factor
 being unity for $\Omega_m=0.3$, $\Omega_{\Lambda}=0.7$ and $h=0.7$,
 and the mass of the merged system $M_t$
 is taken to be close to the total mass of the Galaxy today.
In comparison, the estimated specific energy input from early SNe is
 $\epsilon \sim 0.15 {\rm keV}$ per particle,
 so SF shocks can potentially be more energetic at early Galactic epochs
 and the associated CRs can explain the $^6$Li observations much more naturally \cite{si02}.
Since SF shocks do not eject freshly synthesized CNO nor Fe,
 $\alpha-\alpha$ fusion is the dominant production reaction at low metallicities,
 which can generate large amounts of $^6$Li with little Be or B and no direct correlation with Fe.
A unique, characteristic evolutionary behavior may arise,
 whereby $^6$Li increases very quickly at low metallicity
 (reflecting the main epoch of Galactic SF)
 followed by a plateau or a very slow rise.
This is in marked contrast to models based on SN CRs,
 for which $\log ^6$Li/H vs. [Fe/H] can never be much flatter than linear.
It is also clearly distinct from genuine plateaus
 that may result from exotic production processes in the early Universe \cite{jed00}.

Starting with the SN CR light element evolution model of Suzuki \& Yoshii \cite{sy01},
 we model SF shock CRs in a simple, parameterized way,
 described more fully in \cite{si02}.
In Figure 1,
 we show several SF CR model curves for different parameter values of
 $t_{SF}$, the main epoch of Galactic SF relative to halo chemical evolution,
 $\tau_{SF}$, the main duration of SF,
 and $\gamma_{SF}$, the spectral index of injected particles;
 a standard SN CR model is also displayed.
A CR injection efficiency of 15 \% has been assumed for all cases.
Plotted together are current observational data for $^6$Li \cite{aok02,sln98,nlhd00,asp01}
 (see next section for more discussion),
 along with data for Be \cite{boe99}.
It is apparent that the SN CR model fares well for Be but underproduces $^6$Li,
 whereas the SF CR model is in good agreement with the present $^6$Li data.

\begin{figure}[!htb]
\epsfig{file=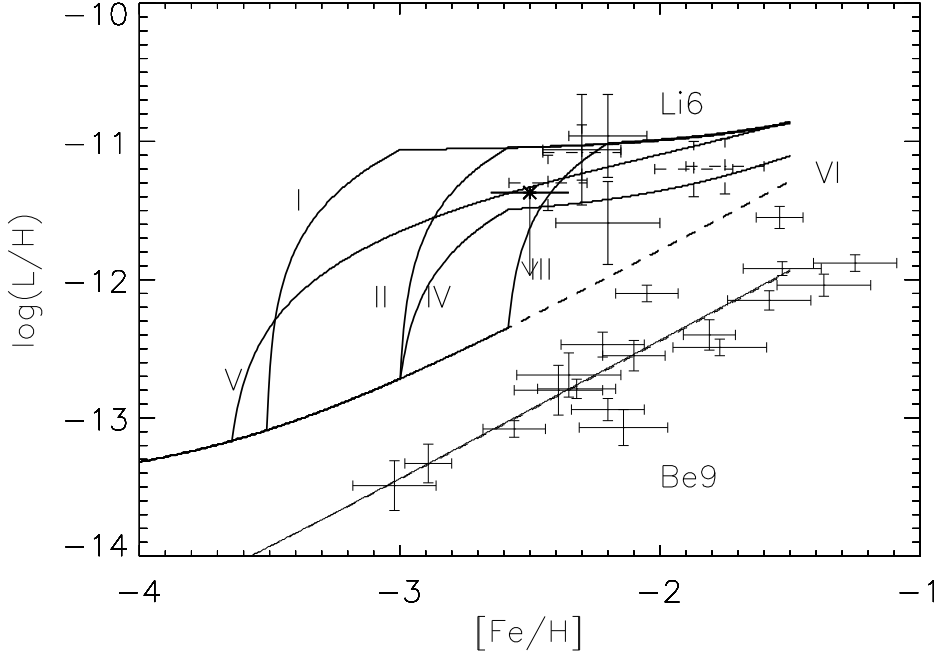,height=9.8cm} 
\caption{
The curves are model results of $^6$Li/H (thick) and Be/H (thin) vs. [Fe/H]
 for SN CRs only (VI, dashed), and SN plus SF CRs (I-V).
The labels correspond to the following sets of parameters
 for $t_{SF}$ [Gyr], $\tau_{SF}$ [Gyr] and $\gamma_{SF}$:
 I (0.12, 0.1, 3), II (0.22, 0.1, 3), III (0.32, 0.1, 3),
 IV (0.22, 0.1, 2) and V (0.1, 0.5, 3).
The markers are current observational data for $^6$Li (thick) and Be (thin),
 the tentative VLT/UVES detections being dashed,
 and the Subaru HDS upper limit for HD140283 being marked with an `x'.
}
\end{figure}

A further, important feature expected in this scenario
 are correlations between the $^6$Li abundance and the kinematic properties of the MPHS.
According to the SF CR picture, $^6$Li production is a direct consequence
 of the principal gas dissipation mechanism of gravitational shock heating, so that
 $^6$Li in MPHS may be interpreted as a fossil record
 of gas dynamical processes during Galaxy formation.
For example, the observed two-component nature of the halo,
 consisting of the inner, flattened and rotating halo 
 and the outer, spherical and non-rotating halo,
 suggest that gaseous dissipative processes have been crucial to the former
 whereas dissipationless stellar dynamics determine the latter \cite{cb00};
 this idea is supported by numerical simulations of Galaxy formation \cite{bc01}.
If this is true, $^6$Li/H should be systematically larger
 in stars belonging to the inner halo compared to those of the outer halo, a testable prediction.
Such correlations are not expected at all in SN CR models.

Thus the SF CR model for $^6$Li has important implications
 for understanding the formation of our Galaxy.
If the above mentioned trends are indeed observed,
 it would not only confirm the SF origin of $^6$Li,
 but may also point to potential new studies of `$^6$Li Galactic archaeology',
 whereby extensive observations of $^6$Li in MPHS can be exploited as unique probes
 of the past dynamical history of our Galaxy.
Some particularly intriguing possibilities include
 testing the hypothesis of non-standard cold dark matter on sub-Galactic scales
 through the early evolution of $^6$Li,
 and probing the efficiency of supernova feedback heating
 through the relative evolution of Be or B and $^6$Li (Inoue \& Suzuki, in prep.).

\section{Recent and future observations of $^6$Li in metal-poor halo stars}

Observing $^6$Li in MPHS is a challenging task,
 as measurement of its weak isotopic shift feature
 relative to the much stronger $^7$Li line requires 
 very high resolution and high S/N spectroscopy \cite{hob00,nis00}.
Previous searches had resulted in positive detections
 for only three stars \cite{sln98,nlhd00} along with a number of upper limits \cite{hob00}.
More recent observations by VLT/UVES
 reveal likely detections for 4 additional stars through a preliminary analysis \cite{asp01}.
A new study by Aoki et al. \cite{aok02}
 discusses the highest quality data yet (S/N $\sim 900-1100$) for $^6$Li in a MPHS,
 that of HD140283 obtained by the Subaru HDS,
 in which analysis with 1D spectrum synthesis models
 leads to a low upper limit on the isotope ratio of $^6$Li/$^7$Li $<0.026$.
These data are all plotted in Figure 1.
Note that although HD140283 is one of the stars tentatively detected by UVES
 as reported in Asplund et al. \cite{asp01},
 the latest analysis of this star by the same authors
 is consistent with the HDS result \cite{aok02}.

Taken at face value, comparison of the HD140283 upper limit with the earlier detections
 suggests either
 a relatively steep increase of $^6$Li/H with metallicity near [Fe/H] $\simeq -2.3$,
 or typical $^6$Li abundances that are a factor of 2--3 lower than the highest measured values.
This can impose interesting constraints on $^6$Li production models,
 provided that stellar depletion effects have not been significant.
For the SN CR models,
 lower observed abundances help in bringing them closer toward agreement,
 but the data are still significantly higher than the conservative SN CR prediction (curve VI;
 see \cite{si02} for details).
A steep log($^6$Li/H)-[Fe/H] relation can arise
 in some secondary CR models depending on the uncertain O-Fe relation \cite{fo99},
 but is unlikely at this abundance level in this metallicity range.
In light of our SF CR model,
 the data set may be consistent with curve IV,
 corresponding either to
 strong shocks with hard CR spectra and less low energy particles for $^6$Li production
 (which is expected if the preshock gas is efficiently cooled by radiation),
 or a total CR energy a factor of $\sim$ 2.7 below the above estimates.
If the steep rise is real, curve III may be a better representation,
 where the main SF shock at the final major merger occurs near [Fe/H] $\sim$ -2,
 possibly being compatible with some other lines of evidence \cite{cb00,bc01}.

The complication with HD140283 is its low surface temperature ($T_{eff} \simeq 5750$ K),
 indicating that stellar depletion of $^6$Li could have been significant.
As the current data are still insufficient to firmly support or rule out different production models,
 high resolution, high S/N spectroscopic observations
 for a large sample of stars with a wide range of metallicities and higher surface temperatures
 are clearly essential.
More detailed and predictive (i.e. less parameterized) modeling of the SF shock scenario,
 using e.g. semi-analytic galaxy modeling techniques (Suzuki, Nagashima \& Inoue, in prep.)
 is also necessary to test its viability more quantitatively.
Such investigations should lead us to decipher the true origin of $^6$Li in MPHS,
 and may also possibly open up a new window on studies of the formation and evolution of the Galaxy.

\end{document}